# Non-demolition Adiabatic Measurement of the Phase Qubit State

G.P. Berman[a], A.A. Chumak[a,b], D.I. Kamenev[a], D. Kinion[c], and V.I. Tsifrinovich[d]

[a]Theoretical Division and CNLS, Los Alamos National Laboratory, Los Alamos, NM 87545, USA

[b]Institute of Physics of the National Academy of Sciences, Pr. Nauki 46, Kiev-28, MSP 03028, Ukraine

[c]Lawrence Livermore National Laboratory, Livermore, CA 94551, USA

[d]Department of Applied Physics, Polytechnic Institute of NYU, 6 MetroTech Center, Brooklyn, NY 11201, USA

**Abstract**

An adiabatic method for a single-shot non-demolition measurement of the phase qubit is suggested. The qubit is inductively coupled to a low-frequency resonator, which in turn is connected with a classical measurement device (phase meter). The resonator drives adiabatic oscillations of the supercurrent in the qubit loop. The back reaction of the qubit loop on the resonator depends on the qubit state. Measuring the phase shift of the resonator's oscillations one can determine the state of the qubit. Numerical computations with available experimental parameters show that the phase difference between the two qubit states increases at a rate of 0.0044 *rad/ns* with the fidelity of about 0.9989 and the measurement time of about 100 ns. The fidelity of the measurement is estimated taking into consideration possible quantum transitions inside and outside the qubit manifold. An increase of the phase difference is possible but it is linked to a reduction of the fidelity. The requirements for the reproducibility of the qubit and resonator parameters are formulated.

1. **Introduction**

The superconducting phase qubit is one of the most promising elements for quantum information processing. (See, for example, [1].) In its simplest version, the phase qubit can be implemented with a superconducting loop interrupted by a single Josephson junction (JJ). Below we will call this the "qubit loop" (QL). The supercurrent in the QL can be described in terms of the phase difference, $\delta$, across the JJ. The QL is biased by a current or external flux so that the potential energy of the QL, $U(\delta)$, can be represented by two wells: the left shallow well and the right deep well (Fig. 1). We will consider a flux-biased phase qubit. The theory of the flux-biased phase qubit is described in detail in [2]. The qubit manifold consists of the two lowest levels corresponding to the wave functions, $\psi_k(\delta)$ $(k=0,1)$, localized in the left shallow well.

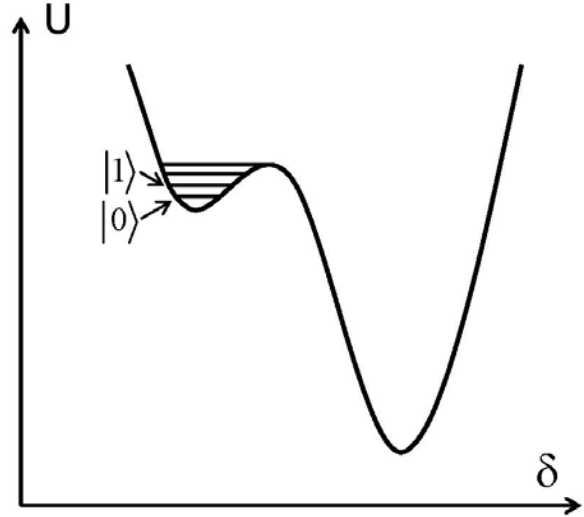

Fig. 1. Potential energy of the QL, and the energy levels in the left shallow well.

The standard scheme for the phase qubit state measurement is the following. Using a microwave pulse, one reduces the height of the barrier separating the two wells. If the qubit is in the first excited state then the QL tunnels from the left shallow well into the deep right well. The supercurrent in the QL as well as the corresponding magnetic flux "jumps" to a new value. This jump can be detected by a nearby SQUID. If the qubit is initially in the ground state the probability of the tunneling is negligible. Thus, measuring the flux produced by the QL supercurrent one can determine the state of the phase qubit. This scheme provides a fast single-shot measurement on the nanosecond time scale with the fidelity slightly below 96% [1]. Note that the qubit manifold consists of the two lowest states in the left well. The measurement described above is a demolition one because in the process of measurement, the QL moves away from the qubit manifold. To the best of our knowledge this single-shot scheme was exploited in all but one existing experiments on the phase qubit measurement (see, for example, [3-6]). The only exception is a multiple-shot measurement of the phase qubit state in experiment [7] where the QL tunneling was detected measuring the frequency shift of a resonator coupled with the QL.

In quantum information processing, a non-demolition projective measurement is preferable (see, for example, [8]). In particular, in the measurement scheme described above the QL tunneling from the qubit manifold reduces the repetition rate of the experiment due to the time required for returning the QL back to the qubit manifold [9]. In this paper we suggest an adiabatic approach for a single-shot non-demolition projective measurement of the phase qubit state using a quasi-classical resonator.

The idea of our approach is the following. We assume that the QL is inductively coupled to the superconducting low-frequency measurement resonator (MR). There is an "adiabatic switch", which allows one to "turn on" or "turn off" the MR-QL coupling adiabatically with respect to the QL and "instantaneously" with respect to the MR. The "adiabatic switch" can be implemented by a variable mutual inductance developed in ref. [10]. In the design of ref. [10] a current bias applied to a DC SQUID controls the screening current, which influences the inductive coupling between two circuits.

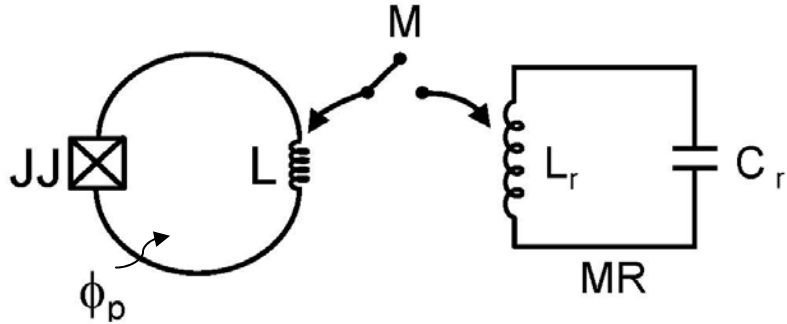

Fig. 2. The measurement scheme suggested in our work. The superconducting loop interrupted by JJ is coupled inductively to a low-frequency resonator. M is the controllable mutual inductance, which can be turned on; $\phi_p$ is the external permanent flux which biases the QL; $L_r$ and $C_r$ are, respectively, the inductance and the capacitance of the MR; $L$ is the QL inductance.

The scheme of our design is shown in Fig. 2. The supercurrent in the MR oscillates with a much lower frequency, than the frequency of the phase qubit. The MR oscillations cause oscillations of the flux in the QL. As a result, the positions of the minima of the potential energy in Fig. 1 adiabatically oscillate. Thus, the supercurrent in the QL adiabatically oscillates with the frequency of the MR. The back action of the QL oscillations on the MR causes a phase shift in the MR oscillations. This phase shift depends on the qubit state. Measuring the phase shift of the MR oscillations, one can determine the state of the phase qubit. Note that the qubit, which is placed initially in one of its basis computational states (ground or excited) remains in the same state during the measurement. If the qubit is placed initially into a superposition of the two basis states, it is expected to collapse quickly to one of its basis states. This should happen because the $z$-component of the phase qubit Bloch vector is an adiabatic invariant of the QL-MR dynamics, which does not change in the process of the phase measurement. Thus, our scheme describes a non-demolition projective measurement of the phase qubit. Note that the interaction between the QL and MR in our scheme is supposed to be strong enough so that the MR phase shift is not negligible in spite of the large difference between the qubit and MR frequencies. The low-frequency oscillations in the resonator are supposed to be amplified with a microstrip SQUID amplifier (MSA) with almost quantum limited noise [11-13]. The phase of the amplified oscillations is measured with a phase meter.

The suggested scheme differs from the "dispersive" multiple-shot non-demolition measurement implemented for the superconducting charge qubits (see, for example, [14]). In the dispersive scheme the energy of the resonator oscillations is small (about one photon in the resonator). The frequency of the resonator is relatively close to the qubit frequency, and the resonator frequency shift is measured using a classical electromagnetic wave interacting with the resonator. Our scheme allows a measurement using the low frequency (below 1 GHz) quasi-classical MR. The low frequency MR might have a significant advantage compared to the high frequency resonators. Indeed, the main source of noise in the dispersive measurement is the noise produced by an amplifier [7,8,15]. At frequencies below 1 GHz the MR oscillations may be amplified by an MSA with almost quantum limited noise. The MSA operating at 500 MHz and 20 mK demonstrated about 25 dB gain with a noise temperature of 50 mK [11] (a single photon at 500 MHz corresponds to 24 mK). Thus, an adiabatic scheme operating at frequencies below 1 GHz may open a way for a single-shot non-demolition measurement of a superconducting qubit. (At higher frequencies the size of the MSA resonator as well as the mutual inductance between the MSA resonator and SQUID decreases resulting in reduction of the gain [11].)

Our scheme differs also from the "adiabatic inversions" suggested in [16]. In the method of adiabatic inversions borrowed from the magnetic resonance force microscopy, the qubit adiabatically oscillates between the ground and the excited state with the frequency of the MR. In this case the qubit motion is adiabatic in the rotating system of coordinates where the Bloch vector retains its direction relative to the effective field. To generate the adiabatic inversions, one has to apply an external resonant field with the Larmor frequency (or a modulated field with a near-resonant frequency), which complicates the measurement scheme. Also the lowest qubit frequency in the rotating system of coordinates is the Rabi frequency, which is much smaller than the Larmor frequency. Thus, during adiabatic inversions, the qubit becomes vulnerable to the low-frequency noise, especially $1/f$ flux noise. In contrary, our scheme described in this work relies on qubit adiabatic evolution in the laboratory system of coordinates, where the only qubit frequency is the Larmor frequency. This scheme does not require an external resonant field because the qubit retains its ground or excited state. The measurement error is expected to be small because the frequencies of the MR as well as $1/f$ noise are small compared to the qubit frequency.

## 2. The Hamiltonian and the equations of motion for the QL-MR system

In this section, we derive the equations of motion for the QL-MR system. Let us denote the supercurrent in the QL as $I$ and the supercurrent in the MR as $I_r$. The total flux through the QL and MR we denote as $\phi$ and $\phi_r$, correspondingly. The total fluxes can be expressed in terms of the supercurrents:

$$\begin{aligned}\phi_r &= MI - L_r I_r, \\ \phi &= \phi_p + MI_r - LI.\end{aligned} \quad (1)$$

Here $M$ is the mutual inductance, $L$ and $L_r$ are the inductances of the QL and MR, $\phi_p$ is the external permanent flux through the QL. From these equations we can express supercurrents in the QL and MR in terms of the fluxes. If $M^2 \neq LL_r$ we have

$$I = (L_r/D)(\phi - \phi_p) - (M/D)\phi_r,$$
$$I_r = (L/D)\phi_r - (M/D)(\phi - \phi_p), \quad (2)$$
$$D = LL_r - M^2.$$

The equations of motion for the MR can be written in terms of the flux, $\phi_r$, and the electric charge on its capacitor, $Q_r$:

$$Q_r/C_r + \dot{\phi}_r = 0,$$
$$\dot{Q}_r = I_r = (L/D)\phi_r - (M/D)(\phi - \phi_p). \quad (3)$$

Here $C_r$ is the MR capacitance. In order to write the equation of motion for the QL we will consider the JJ as a parallel combination of a "pure JJ" and a JJ capacitance, $C$. (See Fig. 3.)

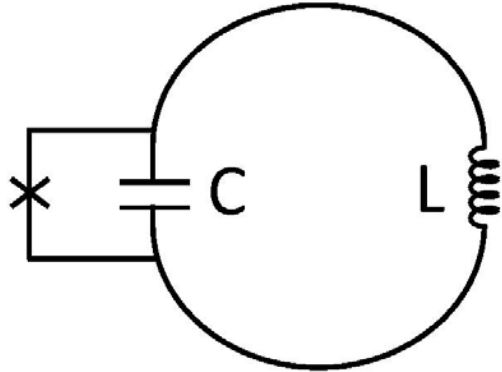

Fig. 3. The QL with the JJ represented as a parallel combination of the JJ capacitance, $C$, and the "pure JJ". "Pure JJ" is indicated by "$\times$".

Now we can write the equations of motion for the QL:

$$\dot{\phi} + Q/C = 0,$$
$$\dot{Q} = I_0 \sin(\phi/\varphi_0) + I. \quad (4)$$

Here $Q$ is the electric charge on the JJ, $\varphi_0 = \hbar/2e$ is the reduced flux quantum, $e$ is the electron charge, $\hbar$ is the reduced Planck's constant, and $I_0$ is the JJ critical current. The system of Eqs. (3)-(4) can be written in the canonical form if we replace our variables (flux and charge) with the canonical variables:

$$Q_r = -p_r / \varphi_0,$$
$$\phi_r = \varphi_0 \varphi_r,$$
$$Q = -p / \varphi_0,$$
$$\phi = \varphi_0 \delta.$$
(5)

The first equation in (5) defines the MR effective "angular momentum," $p_r$, the second equation defines the MR dimensionless flux $\varphi_r$, the third equation defines the QL "angular momentum" $p$, and the last equation defines the phase difference across the JJ, $\delta$. We will also introduce the following parameters:

$$m_r = \varphi_0^2 C_r,$$
$$m = \varphi_0^2 C,$$
$$\lambda_r = L_r / L_0,$$
$$\lambda = L / L_0,$$
$$\mu = M / L_0,$$
$$\Delta = D / L_0^2,$$
$$L_0 = \varphi_0 / I_0,$$
$$E_J = \varphi_0 I_0.$$
(6)

The first two parameters describe the effective "rotational inertia (angular mass)" for the MR and the QL; the next three parameters describe the dimensionless inductances and the mutual inductance; and the last parameter is the Josephson energy. (Where possible, we will use the same notation as in ref. [2].)

Now we can rewrite the equations of motion in the canonical form:

$$\dot{\varphi}_r = p_r / m_r = \partial H / \partial p_r,$$
$$\dot{p}_r = -(E_J / \Delta)[\lambda \varphi_r - \mu(\delta - \varphi_p)] = -\partial H / \partial \varphi_r,$$
$$\dot{\delta} = p / m = \partial H / \partial p,$$
$$\dot{p} = -E_J [\sin \delta + (\lambda_r / \Delta)(\delta - \varphi_p) - (\mu / \Delta)\varphi_r] = -\partial H / \partial \delta,$$
$$H = p_r^2 / 2m_r + p^2 / 2m + E_J \left\{ (\lambda / 2\Delta)\varphi_r^2 + (\lambda_r / 2\Delta)(\delta - \varphi_p)^2 - \cos \delta - (\mu / \Delta)\varphi_r(\delta - \varphi_p) \right\}.$$
(7)

The first two terms in the Hamiltonian, $H$, describe the "kinetic energy" of the MR and QL. The last term is the "potential energy" including the energy of the QL-MR interaction. We will describe the QL quantum mechanically with a wave function, $\psi(\delta, t)$, and the Hamiltonian:

$$H_q = p^2 / 2m + E_J \left\{ (\lambda_r / 2\Delta)(\delta - \varphi_p)^2 - \cos \delta - (\mu / \Delta)\varphi_r \delta \right\},$$
$$p = -i\hbar \partial / \partial \delta.$$
(8)

In this expression, we consider the MR flux $\varphi_r = \varphi_r(t)$ as a slowly (adiabatically) changing parameter.

Next, we will describe MR as a quasi-classical system governed by the Hamiltonian,

$$H_r = p_r^2/2m_r + E_J\left\{(\lambda/2\Delta)\varphi_r^2 - (\mu/\Delta)\varphi_r(\langle\delta\rangle - \varphi_p)\right\}. \tag{9}$$

Here $\langle\delta\rangle$ is the quantum mechanical average of the phase difference across the JJ.

We now solve the coupled system of equations for the QL and MR:

$$\begin{aligned} i\hbar\dot\psi &= H_q\psi, \\ \dot p_r &= -\partial H_r/\partial\varphi_r, \\ \dot\varphi_r &= \partial H_r/\partial p_r. \end{aligned} \tag{10}$$

Here $\psi = \psi(\delta,t)$, $H_q = H_q(p,\delta;\varphi_r)$, $H_r = H_r(p_r,\varphi_r;\langle\delta\rangle)$, $\varphi_r = \varphi_r(t)$, $p_r = p_r(t)$, where $\varphi_r$ is the external parameter for the qubit Hamiltonian, $H_q$, and $\langle\delta\rangle$ is the external parameter for the resonator Hamiltonian, $H_r$. We solve Eqs. (10) in the following way. In the adiabatic approximation, the wave function of the QL is the eigenfunction of the instantaneous Hamiltonian $H_q(\delta,\varphi_r)$. First, we compute this wave function $\psi(\delta,\varphi_r)$ and find the average value $\langle\delta\rangle = \langle\delta(\varphi_r)\rangle$. Next, we substitute this average value into the Hamiltonian, $H_r$, and solve the nonlinear equations for the MR:

$$\begin{aligned} \dot\varphi_r &= p_r/m_r, \\ \dot p_r &= -(E_J/\Delta)\left[\lambda\varphi_r - \mu(\langle\delta\rangle - \varphi_p)\right]. \end{aligned} \tag{11}$$

3. **The MR phase shifts for the two qubit states**

We have taken the following "working values" of the QL parameters from experiment [17]:

$$\begin{aligned} C &= 700\,fF, \\ L &= 720\,pH, \\ I_0 &= 1.7\,\mu A. \end{aligned} \tag{12}$$

The "working values" of the MR parameters and the mutual inductance are chosen as:

$$C_r = 4.4\,pF,$$
$$L_r = 23\,nH, \qquad (13)$$
$$M = 1\,nH.$$

These values are close to those used in experiments with the MSA [12]. The frequency of the unperturbed MR is $f_r = (L_r C_r)^{-1/2}/2\pi = 500\,MHz$.

When the supercurrent in the MR oscillates, the qubit frequency as well as the number of levels in the left shallow well in Fig. 1 adiabatically oscillate. We will assume the following initial conditions for the MR: $\varphi_r(0) = 0$. In our computer simulations we have chosen the value of $p_r(0)$ which corresponds in average to the ten quanta in the MR:

$$p_r^2(0)/2m_r = 10 h f_r,$$

where $h = 2\pi\hbar$. The QL is assumed to be in one of its basis qubit states $|n\rangle$, i.e. the ground state ($n=0$) or the first excited state ($n=1$). We assume the value of the permanent flux, $\varphi_p = 4.992$, which corresponds to the five levels in the left shallow well.

We write the solution of Eqs. (11) for the MR in the form,

$$\varphi_r = \varphi_r(n,t) = A_n \sin\theta_n(t). \qquad (14)$$

Here $A_n$ is the maximum value of the MR dimensionless flux; $\theta_n(t)$ is the phase of the flux oscillations; and the subscript, $n = 0,1$, indicates the initially occupied state of the QL. Time $t$ is counted from the instant of "turning on" the QL-MR coupling. Finally, we compute the phase, $\theta_n(t)$, and the phase shift, $\theta_n(t) - \theta_r(t)$, of the MR flux oscillations with respect to the reference phase $\theta_r(t)$. The reference phase is taken as the phase of the MR flux oscillations with no interaction with the QL (i.e. at $M = 0$). We also compute the phase difference $\theta(t) = \theta_0(t) - \theta_1(t)$, which represents the accuracy of the phase measurement required to distinguish the two qubit states. An increased rate, $\dot\theta$, is associated with a smaller measurement time required to distinguish the two qubit states.

Graphs of the phase shift versus time are shown in Fig. 4 (a)-(c). One can see that, for the "working values" of parameters the phase difference $\theta(t) = \theta_0(t) - \theta_1(t)$ increases at a rate of 0.0044 *rad/ns*. For $M = 1.8\,nH$ and the same values of all other parameters the phase difference increases at a rate of 0.008 *rad/ns*.

Formally, the non-demolition measurement starts when the MR is connected with the classical measurement device – the phase meter. If initially the qubit is in a superpositional state then after the QL-MR coupling the MR will also turn into the superposition of the two quasi-classical coherent states $|\alpha_1\rangle$ and $|\alpha_2\rangle$ with the different rates of the phase accumulation. After the connection to the classical phase meter the qubit collapses to the ground or the first excited state, and the MR collapses to one of its coherent states $|\alpha_1\rangle$ or $|\alpha_2\rangle$ with a definite value of the phase $\arg(\alpha_1)$ or $\arg(\alpha_2)$. The minimum time interval $t_p$ required for measuring the MR phase (relative to the phase of the reference oscillations) can be estimated as a half of the MR period, in our case about $t_p = 1 ns$. Certainly, increasing measurement time above 1 $ns$ would allow one to improve the accuracy of the phase measurement.

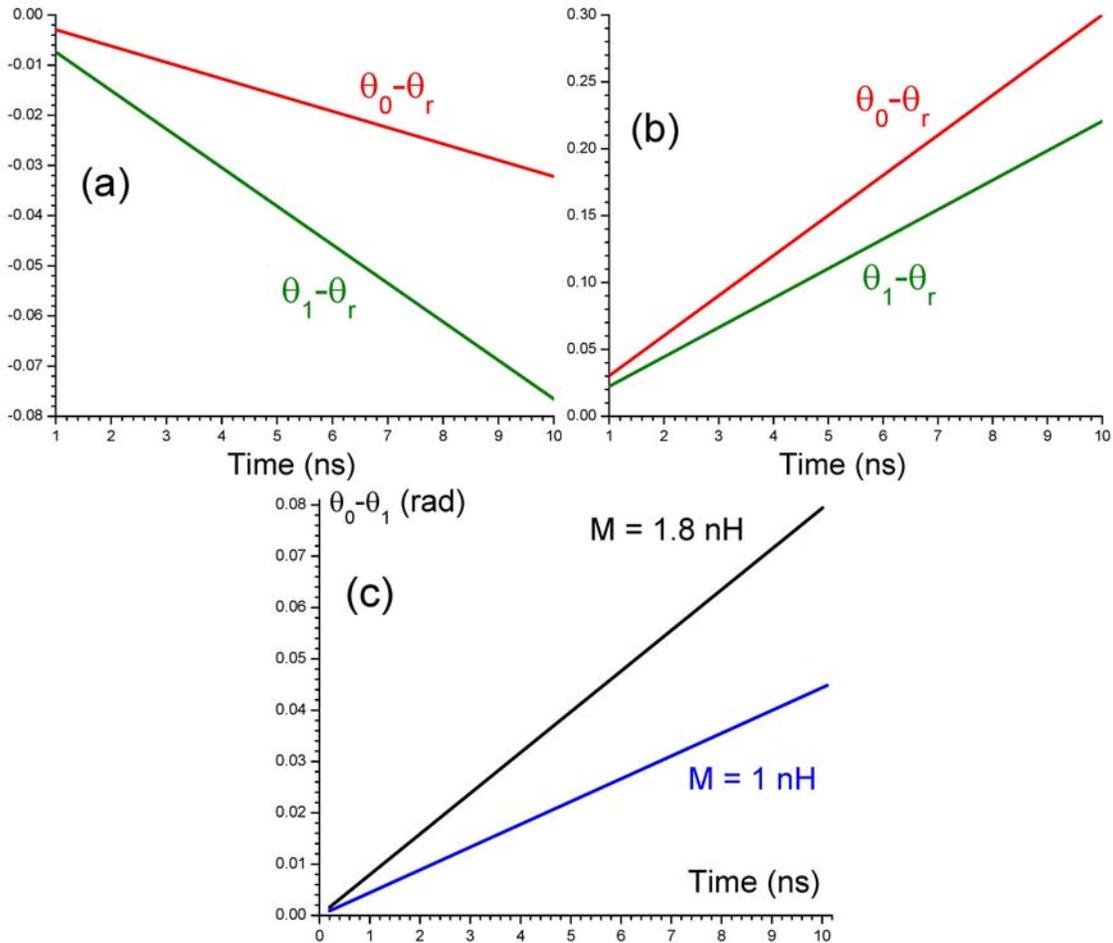

Fig. 4. a) Phase shift $\theta_n(t) - \theta_r(t)$ in radians for the "working values" of parameters (12) and (13), b) the same for $M = 1.8 nH$, c) phase difference $\theta_0(t) - \theta_1(t)$ for two values of $M$.

In fact, we should count the measurement time starting from the "turning on" the QL-MR coupling as during the phase accumulation in the MR one cannot manipulate the qubit (say, apply quantum logic gates). In order to distinguish the two computational states, the MR phase must be measured with accuracy greater than $\theta(t_m)$, where $t_m$ is the time interval between the "turning on" the QL-MR coupling and the beginning of the phase measurement. If we operate with the "working values" of parameters (12) and (13) and start the phase measurement 100*ns* after the QL-MR coupling then the MR phase must be measured with accuracy better than 0.44 rad. The accuracy of the phase measurements reported in literature for electromagnetic oscillations in various systems is better than this value. (As an example, an rf vector voltmeter described in [18] measures phase differences with an accuracy of $1.7 \times 10^{-3} \, rad$.) The MR phase difference must be greater than the quantum phase uncertainty $(4\langle N \rangle)^{-1/2}$, which depends on the average number of quanta $\langle N \rangle$ in the MR [19]. In our simulations $\langle N \rangle = 10$, so that the quantum uncertainty is 0.158 *rad*, which is smaller than the estimated phase difference 0.44 *rad*. Also, the MR phase difference for the two computational states must be greater than the uncertainty of the phase associated with the finite quality factor of the MR. In our case this condition is satisfied. Indeed, even for a low quality factor of about $10^4$, the MR frequency width for the resonant frequency of 500 *MHz* is 50 *kHz*, and the corresponding phase uncertainty in 100*ns* is $\pi \times 10^{-2} \, rad$, which is also smaller than the estimated phase difference.

Finally, we will summarize our scheme of measurement, which includes the following main components: the QL, MR, MSA, and the phase meter. A controllable mutual inductance couples the QL with the MR, the MR with the amplifier, and also the MR with the external coil, which generates oscillations in the MR. At the first step the external coil generates quasi-classical oscillations in the MR. This process is not a part of the measurement but only a preparation to the measurement because during this process the QL is disconnected from the MR. Then the MR is disconnected from the external coil and coupled with the QL. After that during the time interval no less than 100*ns* the MR oscillations accumulate the phase, which depends on the QL state. Then the MR is coupled with the MSA and the phase meter. The phase meter measures the difference between the actual phase of the MR oscillations and the reference phase, which is known from preliminary measurements with no MR-QL coupling. The time of the phase measurement $t_p$ is much smaller than the time of the phase accumulation $t_m$, so we can consider the time $t_m = 100ns$ as estimation for the minimum measurement time.

### 4. Fidelity of the measurement

In the process of measurement, the QL inevitably deviates from a "pure adiabatic trajectory". Consequently, the qubit can "jump" between the two qubit states or even leave the qubit manifold. In either case, the measurement scheme described above will fail. In this section we estimate the fidelity of the measurement.

We will solve the Schrödinger equation for the QL taking into consideration the "instantaneous energy levels", which belong to the left shallow well. We write the solution of the Schrödinger equation in the form:

$$\psi(\delta,t) = \sum_m C_m(t)\psi_m(\delta,t)\exp\left[-\frac{i}{\hbar}\int_0^t E_m(t')dt'\right]. \qquad (15)$$

In this equation $E_m(t)$ and $\psi_m(\delta,t)$ are, respectively, the eigenvalues and eigenfunctions of the instantaneous Hamiltonian, $H_q = H_q(p,\delta;\varphi_r)$, where we substitute $\varphi_r = \varphi_r(n,t)$ found in the adiabatic approximation. [See Eqs. (8) and (14).]

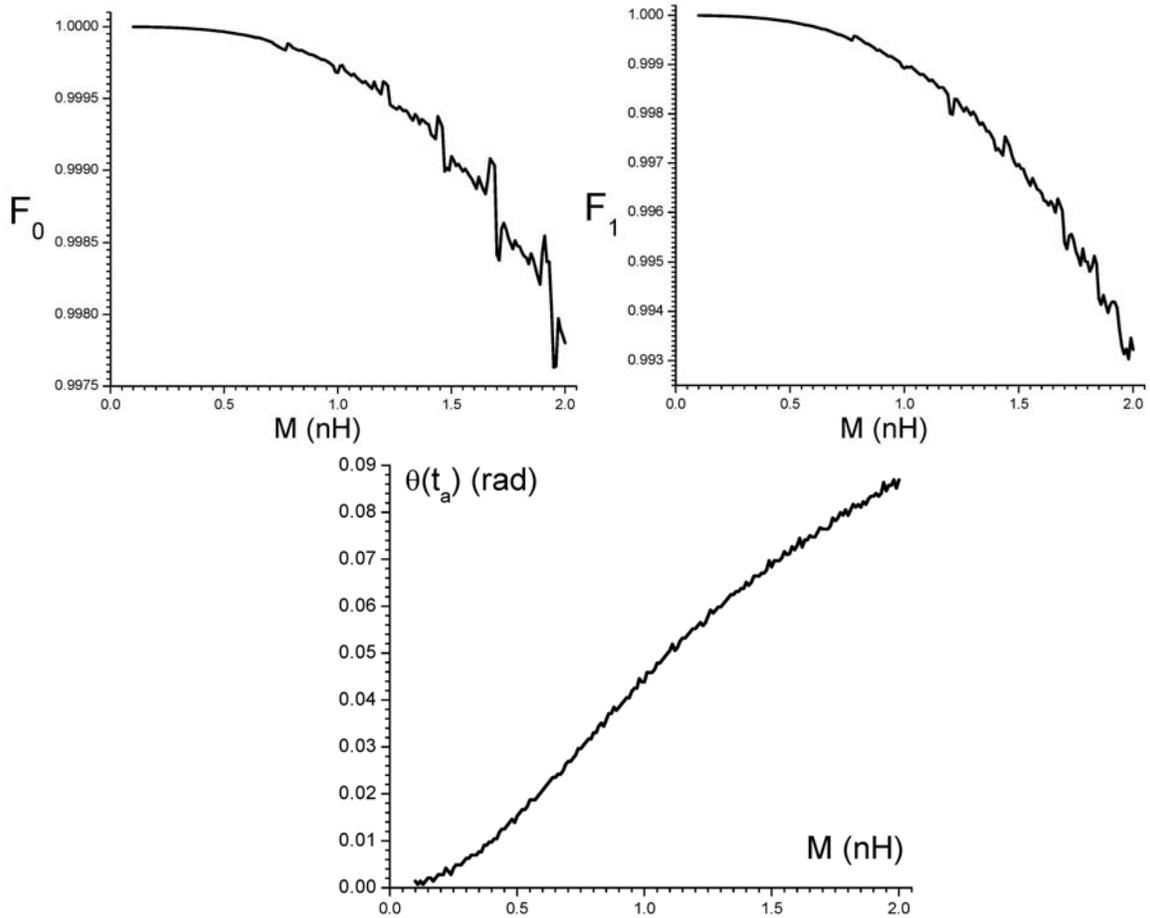

Fig. 5. Fidelity $F_n(n=0,1)$ and phase difference $\theta(t_a)$ vs mutual inductance $M$.

The equation of motion for the amplitudes, $C_k$, is given by

$$\dot{C}_k = \sum_{m\neq k} C_m(t)\exp\left\{(i/\hbar)\int_0^t [E_k(t')-E_m(t')]dt'\right\}\int \psi_k^*(\delta,t)\dot{\psi}_m(\delta,t)d\delta. \qquad (16)$$

We use the initial condition, $C_0(0) = 1$, if the qubit starts from its ground state and $C_1(0) = 1$ if it starts from the first excited state. For all other values of $m$ we take $C_m(0) = 0$. Since the left well depth in Fig. 1 varies, the number of levels, $K$, involved in the dynamics in our computational scheme is also variable and is defined by the condition $|C_{K-1}|^2 \geq 10^{-8}$. If a $k$th level in the left well rises above the barrier separating the two wells, the probability amplitude corresponding to this level is set to zero, $C_k = 0$.

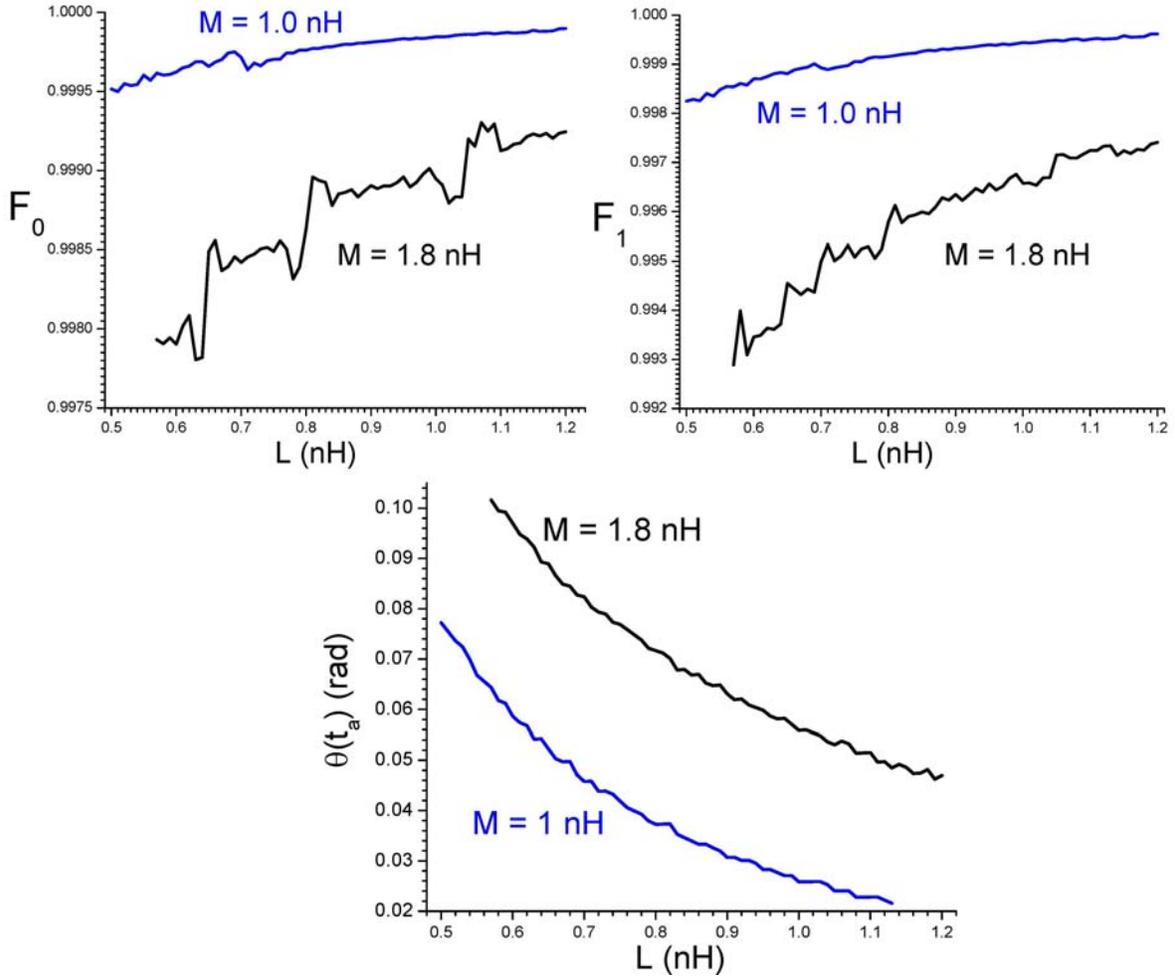

Fig. 6. Fidelity, $F_n (n = 0,1)$, and phase difference, $\theta(t_a)$, vs QL inductance, $L$.

Next, we determine the fidelity of the measurement. We compute the "instantaneous fidelity"

$$f_n(t) = \left| \int \psi_n^*(\delta,t)\psi(\delta,t)d\delta \right|. \tag{17}$$

The "instantaneous fidelity" randomly oscillates with time. In order to obtain the meaningful fidelity of the measurement $F_n$ for the qubit state $|n\rangle$ we average $f_n(t)$ over the averaging time, $t_a$:

$$F_n = \frac{1}{t_a}\int_0^{t_a} f_n(t)dt. \qquad (18)$$

In our computations we take $t_a = 10 ns$, and $n = 0,1$. The fidelity does not change noticeably if the value of $t_a$ increases.

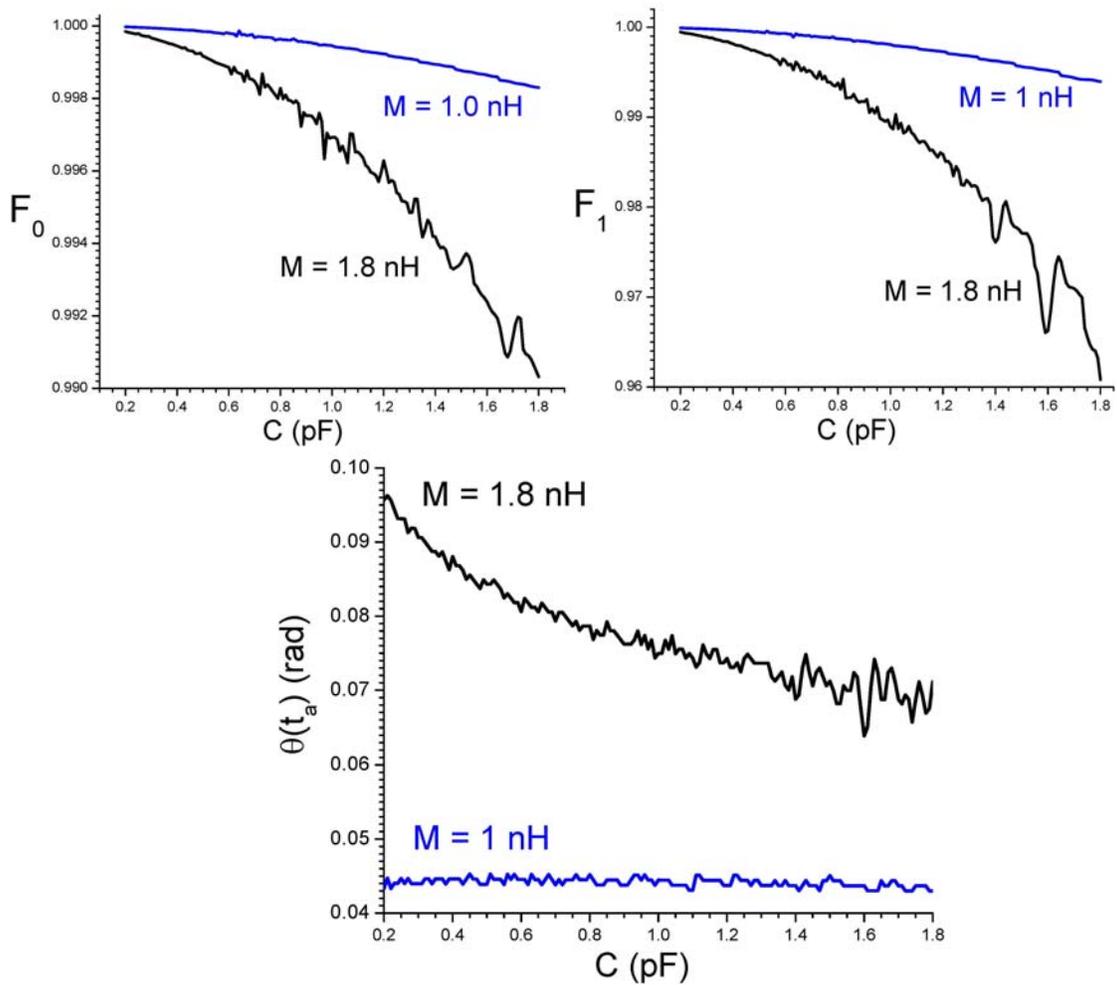

Fig. 7. Fidelity, $F_n (n = 0,1)$, and phase difference, $\theta(t_a)$, vs QL capacitance, $C$.

5. **Results of numerical simulations**

The results of our numerical simulations are presented in Figs 5-10. We show the fidelity, $F_n(n=0,1)$, and the phase difference $\theta(t_a)$ (where $t_a = 10ns$) as functions of the parameters in our simulations. As $\theta(t)$ is proportional to $t$ one can easily find the value of $\theta$ for an arbitrary instant of time. In Figs. 6-10 the two graphs correspond to the two values of the mutual inductance.

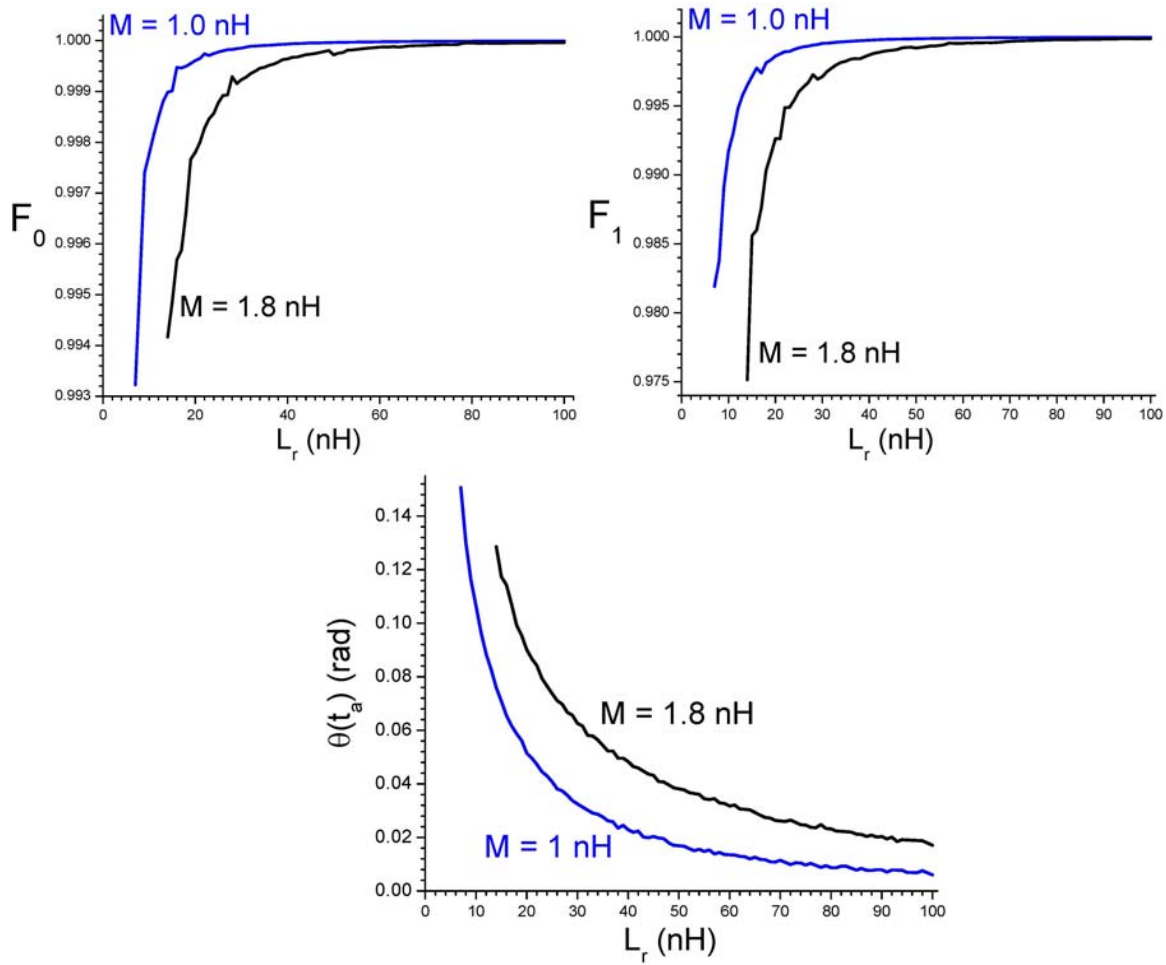

Fig. 8. Fidelity, $F_n(n=0,1)$, and phase difference, $\theta(t_a)$, vs MR inductance, $L_r$.

For all these graphs, the fidelity of the ground qubit state is greater than the fidelity of the excited state. The reason is obvious. If the qubit is in its ground state the probability of error is associated mainly with the unwanted transitions to the first excited state. If the qubit is in its excited state then besides the probability of transition to the ground state it has a comparable probability for the transition to the third level in the shallow well. The jumps and kinks in Figs.

5, 6, and 7 originate from complex dynamics of our quantum multi-level system sensitive to a small change of parameters. Fig. 5 demonstrates the gain of the phase difference and reduction of the fidelity with increasing mutual inductance, $M$. Also, as expected, Figs. 8 and 9 show a sharp decrease in fidelity and an increase of the phase difference when the MR capacitance, $C_r$, or inductance, $L_r$, decrease because in this case the MR frequency becomes closer to the qubit frequency. The graphs in Fig. 6 demonstrate the same tendency: when the QL inductance, $L$, decreases, the phase difference grows, and the measurement fidelity drops. Fig. 7 demonstrates the unique dependences on the QL capacitance, $C$: as $C$ decreases the fidelity grows, and the phase difference either grows or remains the same but does not drop. Thus, a reduction of the QL capacitance is desirable for the adiabatic measurement. Table 1 summarizes the results of our numerical simulations with the different values of the mutual inductance, $M$, and "working values" (12)-(13) of all other parameters.

| M, nH | $N_{min}$ | $N_{max}$ | $f_{min}$, GHz | $f_{max}$, GHz | $F_0$ | $F_1$ | $\theta(t_a)$, rad |
|---|---|---|---|---|---|---|---|
| 0.5 | 4 | 7 | 8.85 | 9.58 | 0.99997 | 0.99987 | 0.020 |
| 1.0 | 4 | 14 | 8.84 | 10.98 | 0.9997 | 0.9989 | 0.044 |
| 1.5 | 4 | 30 | 8.84 | 12.54 | 0.9991 | 0.9970 | 0.070 |
| 1.8 | 4 | 46 | 8.84 | 13.42 | 0.9985 | 0.9950 | 0.080 |
| 2.0 | 4 | 60 | 8.79 | 13.97 | 0.9978 | 0.9932 | 0.086 |

Table 1. Number of levels in the left well, frequency range, fidelities and phase difference for different values of M. $N_{min}$ and $N_{max}$ are the minimum and maximum number of levels in the shallow well in the process of the adiabatic oscillations; $f_{min}$ and $f_{max}$ are the minimum and maximum qubit frequencies; $F_n (n=0,1)$ is the fidelity for the corresponding level; and $\theta(t_a)$ is the phase difference at the end of the averaging time $t = t_a$.

As we mentioned previously, any noise with the frequency much lower than the qubit frequency (in particular $1/f$ noise) is not expected to reduce significantly the fidelity of the adiabatic measurement suggested in this work. However, the spontaneous decay from the first excited state $|1\rangle$ will reduce the fidelity, $F_1$. We will estimate the effect of this spontaneous decay in terms of the Bloch relaxation time, $T_1$. The probability, $P$, of retaining the qubit in the excited state during the time of the phase accumulation can be written as,

$$P = \exp(-t_m / T_1). \tag{19}$$

The fidelity of this measurement can be expressed in terms of this probability: $F_1 = P^{1/2}$. Combining this equation with Eq. (19), we can estimate the minimum relaxation time, $T_1$, which is necessary for a required minimum fidelity, $F_1$: $T_1 = -t_m / 2 \ln F_1$. Table 2 presents the values of $T_1$ found using this expression for $t_m = 100 ns$. The current values of the relaxation time for the phase qubit reported in the literature do not exceed $0.6 \mu s$, and the fidelity for the tunneling measurement is slightly below the value of 0.96 [1].

| $F_1$ | 0.8 | 0.9 | 0.95 | 0.98 | 0.99 | 0.999 | 0.9999 |
|---|---|---|---|---|---|---|---|
| $T_1$ (μs) | 0.22 | 0.47 | 0.97 | 2.5 | 5.0 | 50 | 500 |

Table 2. The minimum relaxation time, $T_1$, which is necessary for the required fidelity, $F_1$.

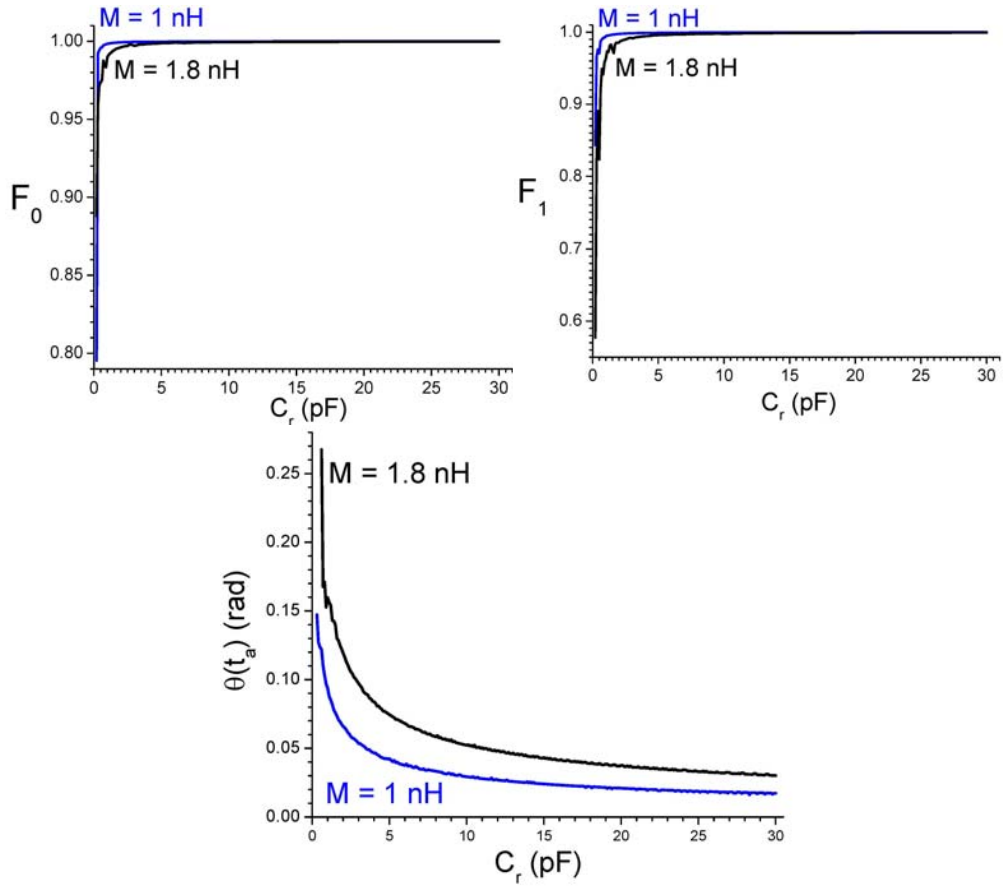

Fig. 9. Fidelity, $F_n (n = 0,1)$, and phase difference, $\theta(t_a)$, vs the MR capacitance, $C_r$.

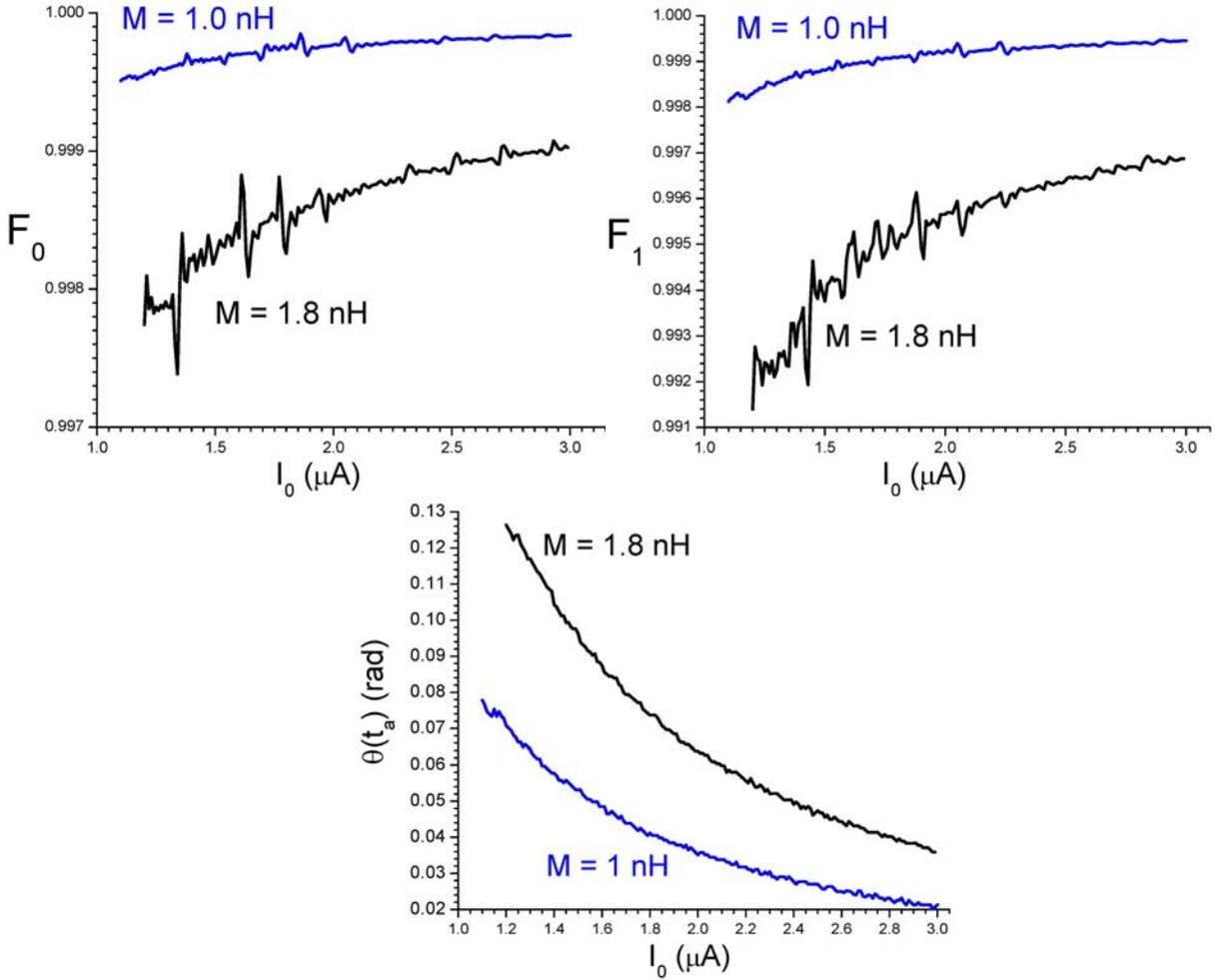

Fig. 10. Fidelity, $F_n (n = 0,1)$, and phase difference, $\theta(t_a)$, vs critical JJ current, $I_0$.

| Parameters | M | L | C | $I_0$ | $L_r$ | $C_r$ |
|---|---|---|---|---|---|---|
| "Working Value" | 1 nH | 0.72 nH | 0.7 pF | 1.7 μA | 23 nH | 4.4 pF |
| Reproducibility Targets | +17% | -22% | +21% | -24% | -13% | -30% |

Table 3. The second row shows the "working values" of the parameters, and the third row represents the reproducibility targets for the required fidelity, $F_1 = 0.9985$. The "+" or "-" sign indicates that fidelity reduces below the minimum required value when the corresponding parameter is greater or smaller than its "working value".

Finally, we will discuss the reproducibility targets for the QL and MR parameters required to produce a given fidelity. Obviously these reproducibility targets depend on the "working values" of the parameters as well as on the required fidelity, $F_1$. (We can take into consideration only the fidelity found for the excited state because $F_0 > F_1$.) As an example we consider a case in which the "working values" of parameters are given by Eqs. (12)-(13) and the required fidelity is $F_1$=0.9985. It follows from Table 1 that for the "working values" of parameters, $F_1 = 0.9989$. The question arises: what minimum deviation in the values of parameters will reduce the measurement fidelity below its required value? These minimum deviations can be considered as reproducibility targets for the corresponding parameters. In our example, the reproducibility targets are presented in the third row of Table 3. This Table shows that the measurement fidelity will reduce below the required value, $F_1 = 0.9985$, when the mutual inductance increases by more than 17% from its "working value", $M = 1nH$, or the qubit inductance decreases by more than 22% from its "working value", $L = 0.72nH$, and so on. One can see that the most sensitive (critical) parameters for the measurement fidelity are the MR inductance, $L_r$, and the mutual inductance, $M$: the reproducibility targets for these parameters are below 20%.

## 6. Conclusion

We have suggested an adiabatic method for a single-shot non-demolition measurement of a state of the phase qubit. In our method, the low-frequency MR induces adiabatic oscillations of the supercurrent in the QL, which is associated with the average phase difference across the JJ. The adiabatic oscillations of the supercurrent in the QL cause a back reaction on the MR - the MR phase shift, which depends on the state of the qubit. This allows the measurement of the qubit state. The expected advantages of our method are: a) a high fidelity of the measurement, and b) the opportunity of using a low-frequency MSA, which demonstrated the high gain and almost quantum limited noise. The high fidelity is associated with the measurement technique: this method does not require the microwave pulses, and the qubit remains in its initial (ground or excited) state in the laboratory system of coordinates. Using realistic parameters (12)-(13) we have obtained the rate of increase of the MR phase difference 0.0044 *rad/ns* with the fidelity of 0.9989 and the measurement time of about 100 ns. The phase difference can be enlarged by increasing the MR frequency or the mutual inductance but this is linked to a reduction of the measurement fidelity. We also suggested a method of computing the reproducibility targets for the QL and MR parameters.

**Acknowledgement**

This work was carried out under the auspices of the National Nuclear Security Administration of the U.S. Department of Energy at Los Alamos National Laboratory under Contract No. DE-AC52-06NA25396 and by Lawrence Livermore National Laboratory under Contract DE-AC52-07NA27344, and was funded by the Office of the Director of National Intelligence (ODNI), and Intelligence Advanced Research Projects Activity (IARPA). All statements of fact, opinion or conclusions contained herein are those of the authors and should not be construed as representing the official views or policies of IARPA, the ODNI, or the U.S. Government.